\documentclass[letterpaper,notitlepage,nofootinbib,superscriptaddress]{revtex4-1} 

\DeclareFontFamily{OT1}{pzc}{}
\DeclareFontShape{OT1}{pzc}{m}{it}{<-> s * [1.10] pzcmi7t}{}
\DeclareMathAlphabet{\mathpzc}{OT1}{pzc}{m}{it}

\usepackage[utf8]{inputenc}
\usepackage{amsmath,amssymb,amsfonts}
\usepackage{graphicx,wrapfig}
\usepackage{dcolumn}
\usepackage{bm}
\usepackage{color}
\usepackage{bbold}
\usepackage{url}
\usepackage{slashed}
\usepackage[svgnames]{xcolor}
\usepackage[english]{babel}
\usepackage{microtype}

\usepackage[colorlinks=true,linkcolor=blue,urlcolor=blue,citecolor=blue]{hyperref}

\usepackage{color}

\definecolor{darkgreen}{rgb}{0,0.65,0}

\newcommand{\be}{\begin{eqnarray}}
\newcommand{\ee}{\end{eqnarray}}
\newcommand{\ba}{\begin{array}}
\newcommand{\ea}{\end{array}}

\begin{document}

\title{Forces inside the nucleon on the light front from 3D Breit frame force distributions:
Abel tomography case}
\author{Julia Yu.~Panteleeva}
\affiliation{Ruhr University Bochum, Faculty of Physics and Astronomy,
Institute for Theoretical Physics II, D-44870 Bochum, Germany}
\author{Maxim V.~Polyakov}
\affiliation{Ruhr University Bochum, Faculty of Physics and Astronomy,
Institute for Theoretical Physics II, D-44870 Bochum, Germany}
\affiliation{Petersburg Nuclear Physics Institute, 
		Gatchina, 188300, St.~Petersburg, Russia}

\begin{abstract}
We derive simple relations which express 2D light front force distributions in terms of 3D Breit frame pressure and shear force distributions.
Mathematically the relations correspond to invertible Abel transformation and they establish one-to-one mathematical equivalence of
3D Breit frame force distributions and 2D light front ones.

Any knowledge  (model calculation, experimental measurement, etc.) about pressure and shear force distributions  in Breit frame can be unambiguously  transformed into
light front force distributions with the help of Abel transformation. 
It is important that the transformation ensures 2D stability conditions if the 3D stability conditions are satisfied. 

As an illustration of how the relations work, we calculated the light front force distributions for a large nucleus as a liquid drop, and for 
large $N_c$ nucleon as a chiral soliton. 

\end{abstract}

\maketitle


\section*{Introduction}
\label{sec:Introduction}

The linear response of a hadron to a  change of the external space-time metric is described by the gravitational form factors (GFFs). 
The GFFs contain rich information about the internal structure of hadrons, for a detailed review see, e.g.  
Ref.~\cite{Polyakov:2018zvc}. Particular interest for us here is the mechanical properties -- elastic pressure and shear force distributions
inside the hadron of a mass $m$. These distributions can be characterised by  the Breit frame (BF) static energy momentum tensor (EMT)
  \cite{Polyakov:2002yz}
 \footnote{ Although it is not necessary for derivations here, we note that
 for the distances $r\gg1/(2m)$ the Breit frame static EMT has clear probabilistic interpretation,
for the distances $\sim 1/(2m)$ it can be viewed  as quasi-probabilistic phase-space average (in Wigner sense) of the observables
in the rest frame of the hadron.
 Detailed discussion of this issue
can be found in Ref.~\cite{Lorce:2020onh}. We stress that the interpretation of the BF static EMT is not relevant for the derivation here.}:
\be
\label{eq:BFEMT}
\Theta^{\mu\nu} (\vec r ) 
= 
\int {d^3 \Delta \over  (2\pi)^3 2E } e^{-i \vec \Delta \cdot \vec r} 
\langle p^\prime \, |{\hat \Theta}_{\rm QCD}^{\mu\nu}(0)|p \rangle .
\ee
Here ${\hat \Theta}_{\rm QCD}^{\mu\nu}(0)$ is the QCD energy-momentum tensor (EMT) operator the matrix element of which is computed between hadron states with  
momenta $p^0=p^{0\prime}=E=\sqrt{m^2+\vec \Delta^2/4}$,  and $p^{i\prime}=-p^i=\Delta^i/2$. The $00$ component of the static EMT contains the information
about the energy distribution inside the hadron, $0i$ component about the spin distribution, and $ik$ components provide us the 
distribution of elastic pressure and shear
forces inside the hadron  \cite{Polyakov:2002yz}. 

In relativistic quantum field theory it is impossible to localise one-particle state with accuracy better than its Compton wave length $\lambdabar= \frac{\hbar}{m c}$.
Therefore, at the distances smaller or of the order of $\lambdabar$ one has to interpret the Breit frame static 
EMT (\ref{eq:BFEMT}) from a qasi-probabilistic phase-space perspective \cite{Lorce:2018egm,Lorce:2020onh}.
For the first time the phase-space picture to connect BF to light front was in the study of the angular momentum distribution \cite{Lorce:2017wkb}.
The static EMT (\ref{eq:BFEMT}) can be viewed as  the Wigner phase-space average \cite{Wigner:1932eb,Hillery:1983ms}
of
the force distributions inside the rest frame nucleon. See detailed discussion in Ref.~\cite{Lorce:2018egm} for EMT
and Ref.~\cite{Lorce:2020onh} for the charge densities. Due to Heisenberg's uncertainty principle the Wigner distributions have only
quasi-probabilistic interpretation. In the non-relativistic limit (which corresponds to distances $r\gg \frac{1}{2m}$ in Eq.~(\ref{eq:BFEMT})) the Wigner distributions
acquire strict probabilistic interpretation, see detailed discussion in  \cite{Lorce:2018egm,Lorce:2020onh,Lorce:2017wkb}.

  If one insists on strict probabilistic
interpretation of distributions, the static EMT (\ref{eq:BFEMT}) for $r\sim 1/(2 m)$ acquires the so-called relativistic corrections  discussed since 1950th \cite{yennie},
for more recent discussion see Refs.~\cite{Burkardt:2000za,Miller:2010nz,Jaffe:2020ebz}.
The relativistic 
corrections can be kinematically suppressed if one considers the distributions  in the infinite momentum frame (IMF) or if one uses
quantization on the light front, 
see e.g. Ref~\cite{Burkardt:2000za,Miller:2010nz}.
In recent paper \cite{Lorce:2020onh}, using the phase-space Wigner distributions,
the natural interpolation between the Breit frame and IMF charge distributions was obtained, such analysis can be repeated
for the case of force distributions. 

The densities for the internal force distributions in the nucleon in IMF and on light front were derived first in Ref.~\cite{Lorce:2018egm} 
with help of Wigner phase-space distribution. More recently, light front force distributions were also obtained in 
Ref.~\cite{Freese:2021czn} using methods of the light-cone quantization. 
These densities have strict probabilistic interpretation (no relativistic corrections)
and they are identical in both approaches. The corresponding densities are two-dimensional (2D) and are distorted relative 3D quasi-probabilistic
distributions due to the motion of the nucleon relative to observer. 

Our aim here is to relate the 2D light front densities of internal forces inside the hadron to the 3D 
Breit frame distributions. We derive one-to-one relations between light front and Breit frame distributions:
any result (model calculation, experimental measurement, etc.) for 3D Breit frame forces can be unambiguously
translated to the form of the 2D light front force distributions (and vice versa). 
Additionally the derivation of the 2D stability conditions from those of 3D are given.
The relations are illustrated on examples of 1) heavy nucleus as a liquid drop, and 2) nucleon at large $N_c$.

Working with the 3D Breit frame distributions we can use our ``rest frame" physics intuition, and if one is not
comfortable with their quasi-probabilistic interpretation at $r\sim \lambdabar$, one can always use the relations
derived here
to transform  unambiguously these distributions to the light front ones and vice versa.

\section*{Relations between 3D Breit frame and 2D light-front force distributions}

For the nucleon there are three independent EMT form factors 
\cite{Kobzarev:1962wt,Pagels:1966zza}:\footnote{For definition of the gravitational form factors for hadrons of arbitrary spin see Ref.~\cite{Cotogno:2019vjb}.}
\begin{eqnarray}
\langle p'|\hat \Theta^{\mu\nu}_{\rm QCD}(0)|p\rangle &=& \bar u(p') \left[  A(t) \frac{ P^\mu P^\nu}{m} +   J(t)
\frac{i  P^{\{\mu} \sigma^{\nu \}\alpha} \Delta_\alpha }{m}
+ \frac {D(t)}{4 m}  \left(\Delta^\mu \Delta^\nu-\eta^{\mu\nu} \Delta^2\right) \right]  u(p) \, ,
\label{EMTdef}
\end{eqnarray}
where $\hat \Theta^{\mu\nu}_{\rm QCD}(x)$ is the symmetric EMT operator of QCD,
  $P=(p+p')/2$, $\Delta=p'-p$,  $t=\Delta^2$, and symmetrisation operation 
     is defined as
$X_{\{\mu} Y_{\nu\}}=\frac 12 (X_\mu Y_\nu+X_\nu Y_\mu)$. The values of the nucleon EMT form factors at zero momentum transfer provide us with three basic mechanical characteristics of the nucleon --
the mass $m$, the spin $J=1/2$, and the D(ruck)-term $D(0)$. While the mass and spin of the nucleon are well-studied and well-measured quantities, the third mechanical characteristics -- the {\it Druck} term or $D$-term--
is more subtle, as it is related to the distribution of the internal  forces inside the nucleon \cite{Polyakov:2002yz}. Important distinguishing
feature of the nucleon $D$-term is that to access it one needs variations of the space-time metric such that the resulting Riemann tensor is non-zero.  
This feature of the Druck term is especially clearly seen in terms of effective field theory for pions and nucleon in curved space
discussed recently in details in Ref.~\cite{Alharazin:2020yjv}.
To access the mass and the spin it is enough to
perform the   variation of the metric  with zero Riemann tensor, e.g. 
  to go to a non-inertial coordinate system.
Despite of difficulty to access the Druck form  factor, the first experimental data are available for the nucleon 
  \cite{Nature,Kumericki:2019ddg,Dutrieux:2021nlz} and for the pion \cite{Kumano:2017lhr}. However the systematic and statistical uncertainties
  are still large. This will be considerably improved with new experiments on hard exclusive processes.

  The  Breit frame distributions of the elastic pressure $p(r)$ and shear force $s(r)$  in 3D can be obtained in terms of Druck form factor $D(t)$
  through \cite{Polyakov:2002yz,Polyakov:2018zvc}:
\begin{eqnarray}
\label{Eq:relationSPD}
	s(r)= -\frac{1}{4 m}\ r \frac{d}{dr} \frac{1}{r} \frac{d}{dr}
	{\widetilde{D}(r)}, \quad
	p(r)=\frac{1}{6 m} \frac{1}{r^2}\frac{d}{dr} r^2\frac{d}{dr}
 	{\widetilde{D}(r)}, \quad
	{\widetilde{D}(r)=}
	\int {\frac{d^3{\bf  \Delta}}{(2\pi)^3}}\ e^{{-i} {\bf  \Delta r}}\ D(-{\bf  \Delta}^2).
\end{eqnarray}
Recently in Refs.~\cite{Lorce:2018egm,Freese:2021czn} the 2D light front
pressure and shear force distributions were obtained in terms of the same Druck form factor $D(t)$ (we follow closely the notations of Ref.~\cite{Freese:2021czn}.):
\be
\widetilde{D}(x_\perp)=\dfrac{1}{4P^+}\int\dfrac{d^2\mathbf{\Delta}_\perp}{(2\pi)^2}D(-\mathbf{\Delta}^2_\perp)e^{-i\mathbf{\Delta}_\perp\cdot \mathbf{x}_\perp},
\ee
\be
\label{eq:p2ds2d}
p^{(2D)}(x_\perp)=\dfrac{1}{2x_\perp}\dfrac{d}{dx_\perp}\left(x_\perp\dfrac{d}{dx_\perp}\widetilde{D}(x_\perp)\right),\hspace{15pt}
s^{(2D)}(x_\perp)=-x_\perp\dfrac{d}{dx_\perp}\left(\dfrac{1}{x_\perp}\dfrac{d}{dx_\perp}\widetilde{D}(x_\perp)\right),
\ee
where $\mathbf{x}_\perp$ is the 2D vector in the transverse plane. Identical up to a (conventional) global rescaling by $P^+/m$ equations were derived in Ref.~\cite{Lorce:2018egm}, using Wigner phase-space distribution in IMF.

As the 2D and 3D force distributions are expressed in terms of the same Druck form factor $D(t)$, they can be easily related to each other.
Below we give these relations. For convenience we introduce 2D pressure  $\mathpzc{P}(x_\perp)$ and shear force distribution $\mathpzc{S}(x_\perp)$
which differ from those in Eq.~(\ref{eq:p2ds2d}) by multiplication with the Lorentz factor $\dfrac{P^+}{2m}$:

\be
\mathpzc{S}(x_\perp)=\dfrac{P^+}{2m}s^{(2D)}(x_\perp),\hspace{15pt} \mathpzc{P}(x_\perp)=\dfrac{P^+}{2m}p^{(2D)}(x_\perp).
\ee
These 2D force distributions can be easily obtained in terms of 3D distributions defined in Eq.~(\ref{Eq:relationSPD}):

\be
\label{eq:p2ds2dr}
\mathpzc{S}(x_\perp)=\int\limits_{x_\perp}^{\infty} \dfrac{dr}{r} s(r) \dfrac{  x^2_{\perp}}{\sqrt{r^2-x^2_{\perp}}}, \hspace{15pt}
 \dfrac{1}{2}\mathpzc{S}(x_\perp)+\mathpzc{P}(x_\perp)=\int\limits_{x_\perp}^{\infty} \dfrac{dr}{r}  s(r) \sqrt{r^2-x^2_{\perp}}.
\ee
These equations have the form of invertible Abel transformation \cite{Abel}. 
The functions $\mathpzc{S}(x_\perp)/x_\perp^2$ is the Abel image of $s(r)$.
For the readers' convenience we collected basic facts about the Abel transformation in the Appendix.

The Eqs.~(\ref{eq:p2ds2dr}) can be obtained from analogous equations in Ref.~\cite{Lorce:2018egm} by change of
  of the integration variable. Here we derive new inversion equations.
The inverse to (\ref{eq:p2ds2dr}) transformation can be easily obtained:

\be
\label{eq:p3ds3dr}
s(r)=-\dfrac{2}{\pi} r^2 \int\limits_r^\infty dx_\perp \dfrac{d}{d x_\perp}\left(\dfrac{\mathpzc{S}(x_\perp)}{x_\perp^2}\right) \dfrac{1}{\sqrt{x_\perp^2-r^2}}, \quad
\frac 23 s(r)+p(r)=\frac 4\pi \int\limits_r^\infty \dfrac{dx_\perp}{x_\perp}\mathpzc{S}(x_\perp) \dfrac{1}{\sqrt{x_\perp^2-r^2}}.
\ee
We obtain very interesting result: the normal force distribution in 3D ($\frac 23 s(r)+p(r)$) is the Abel image of the 
light front shear force distribution $\frac 4\pi\mathpzc{S}(x_\perp) $ in 2D. Both Eq.~(\ref{eq:p2ds2dr}) and Eq.~(\ref{eq:p3ds3dr}) 
can be compactly written as:

\be
\label{eq7:abelform}
\mathpzc{S}(x_\perp)=x_\perp^2\ A[s](x_\perp), \quad 
\frac 12 \mathpzc{S}(x_\perp)+\mathpzc{P}(x_\perp)=\frac 12 A\left[r^2 \left(\frac 23 s+p\right)\right](x_\perp)\\
s(r)=-\frac{2}{\pi} r^2 A\left[x_\perp \dfrac{d}{d x_\perp}\left(\dfrac{\mathpzc{S}(x_\perp)}{x_\perp^2}\right)\right](r),       
       \quad \frac 23 s(r)+p(r)=\frac 4\pi\ A[\mathpzc{S}](r).
\label{eq8:abelform}
\ee
See Appendix for the explanation of the Abel transformation operation $A[...](...)$.

From Eq.~(\ref{eq:p2ds2dr}) it is easy to check that the 2D von Laue stability condition for the pressure: 
\be
\int d^2 \mathbf{x}_\perp \mathpzc{P}(x_\perp)=0
\ee 
is satisfied automatically. This 2D stability condition was discussed for the first time in section 5.1.2 of \cite{Lorce:2018egm}.
Also it is easy to see that
 $\mathpzc{S}$ and $\mathpzc{P}$ satisfy automatically the EMT conservation differential equation:
\be
\mathpzc{P}'(x_\perp)+\dfrac{1}{2}\mathpzc{S}'(x_\perp)+\dfrac{1}{x_\perp}\mathpzc{S}(x_\perp)=0.
\ee 
This equation implies a number of integral relations between $\mathpzc{S}(x_\perp)$ and $\mathpzc{P}(x_\perp)$, in particular
it is easy to show that:

\be
\int\limits_0^{\infty} d {x}_\perp \left(-\frac 12 \mathpzc{S}(x_\perp)+\mathpzc{P}(x_\perp)\right)=0.
\ee
The combination $-\frac 12 \mathpzc{S}(x_\perp)+\mathpzc{P}(x_\perp)$ has the meaning of the transverse 
force distribution (eigenvalue of 2D stress tensor). From above equation we can conclude that this distribution 
must have at least one node. Number of other integral relations between 2D force distributions can be easily derived
using the general method described in Appendix~A of Ref.~\cite{Polyakov:2018zvc}.

The Druck-term can be expressed in terms of 2D force distributions as:
\be
D(0)=-m \int d^2x_\perp x^2_\perp\mathpzc{S}(x_\perp)=4 m \int d^2x_\perp x^2_\perp\mathpzc{P}(x_\perp).
\ee
Surely the resulting Druck-term $D(0)$ coincides with that obtained from 3D Breit frame force distributions:

\be
D(0)=-\frac{4}{15} m\int d^3r\ r^2\ s(r)=m\int d^3r\ r^2\ p(r).
\ee


\subsection*{Local stability conditions}

In Refs.~\cite{Polyakov:2018zvc,Perevalova:2016dln,Lorce:2018egm} it was  argued that for the stability of the mechanical system the 3D
pressure and shear forces should satisfy the following inequality:
\be
\label{eq:stab1}
\frac{2}{3} \, s(r) +p(r) > 0\,.
\ee
From Eq.~(\ref{eq7:abelform}) we see immediately that this 3D stability condition implies:
\be
\label{eq:stab2}
 \dfrac{1}{2}\mathpzc{S}(x_\perp)+\mathpzc{P}(x_\perp)>0,
\ee
as the Abel image of a positive function is also positive.
The above inequality is analogous to the 3D local stability conditions (\ref{eq:stab1}) because
$\frac{1}{2}\mathpzc{S}(x_\perp)+\mathpzc{P}(x_\perp)$ corresponds to the distribution of normal force in 2D.
This local stability condition was discussed in Ref. \cite{Freese:2021czn}. 
We think it is important result that the stability conditions in 3D imply the stability of 2D mechanical system.

One can show that the 3D stability condition (\ref{eq:stab1}) can be obtained from the positivity of the shear force distribution (pressure anisotropy) $s(r)>0$.
Indeed, from the 3D EMT conservation equation:
\be
\dfrac{d}{dr} \left(\frac{2}{3} \, s(r) +p(r)\right) =-\dfrac{2}{r} s(r),
\ee
for $s(r)>0$ we conclude that $\frac{2}{3} \, s(r) +p(r)$ is a monotonically decreasing function. As this function goes to zero it must be positive.
From Eq.~(\ref{eq:p2ds2dr})  we see immediately that the condition $s(r)>0$  implies that:

\be
\mathpzc{S}(x_\perp)>0, \hspace{15pt} \dfrac{1}{2}\mathpzc{S}(x_\perp)+\mathpzc{P}(x_\perp)>0.
\ee
So we see that the stronger 3D stability conditions $s(r)>0$ implies also analogous  stronger condition in 2D. 

We note that to ensure the 3D stability condition (\ref{eq:stab1}) we need stronger stability condition
in 2D $\mathpzc{S}(x_\perp)>0$, see Eq.~(\ref{eq8:abelform}). The condition  (\ref{eq:stab2}) is not enough to guarantee the stability in 3D.

Using the positivity of the normal forces we can introduce the transverse mechanical radius: 
\be
\label{eq:mechr}
\langle x_\perp^2\rangle_{\text{mech}}=\dfrac{\int d^2x_\perp x^2_\perp\left( \dfrac{1}{2}\mathpzc{S}(x_\perp)+\mathpzc{P}(x_\perp)\right)}{\int d^2x_\perp \left( \dfrac{1}{2}\mathpzc{S}(x_\perp)+\mathpzc{P}(x_\perp)\right)}=\dfrac{4D(0)}{\int\limits_{-\infty}^0 dt D(t)}=\dfrac{2}{3}  \langle r^2\rangle_{\text{mech}}.
\ee
Our result, which is two times smaller then the result of Ref.~\cite{Freese:2021czn}, corresponds to usual geometric ratio of 2/3 between 2D and 3D 
mean square radii. For derivation of the relation between 2D and 3D radii the following relation for Mellin moments of Abel images can be used:

\be
\int\limits_0^\infty dx_\perp x_\perp^{N} A[g](x_\perp)=\frac{\sqrt{\pi}}{2}\  \dfrac{\Gamma\left(\frac{N+1}{2}\right)}{ \Gamma\left(
\frac{N+2}{2}\right)} \int\limits_0^\infty dr\ r^{N-1} g(r).
\ee

\subsection*{2D pressure at the origin and the line tension}
The pressure in the centre of the nucleon can be easily obtained from Eq.~(\ref{eq:p2ds2dr})\footnote{In our derivation we assume that 3D $s(r)$ is not singular at $r=0$,
which corresponds to nullification of 2D $\mathpzc{S}(0)$.}:
\be
\mathpzc{P}(0)=\int\limits_0^{\infty} dr\ s(r) =\gamma^{(3D)}.
\ee
Here $\gamma^{(3D)}$ has the meaning of the surface tension of the 3D mechanical system. 
Obviously the 2D pressure in the centre satisfies Kelvin relation, generalised to 2D:
\be
\mathpzc{P}(0)=\int\limits_0^{\infty} \dfrac{dx_\perp}{x_\perp} \mathpzc{S}(x_\perp),
\ee
this relation was derived for the first time in Ref.~\cite {Lorce:2018egm}.
Using the properties of the Abel transformation one can easily see that for the case of a singular normal force distribution in 3D
the corresponding singularity of 2D counterpart is softer by one unit, for example $1/r \to \ln(1/x_\perp)$, $1/r^2 \to 1/x_\perp$, etc.

The 2D line tension (2D counterpart of the surface tension in 3D) $\gamma^{(2D)}$ can be obtained as:

\be
\label{eq:gamma2}
\gamma^{(2D)}= \int\limits_0^\infty dx_\perp \mathpzc{S}(x_\perp)=\frac{\pi}{4} \int\limits_0^\infty dr\ r\ s(r).
\ee

\section*{\boldmath Light-front force distributions in a liquid drop}

We start with illustration of the 3D-2D relations from previous section on the example of a very simple
mechanical system -- the liquid drop.
We hope that consideration of such simple mechanical system on the light front allows us to develop 
physics intuition about 2D force distributions.

For a large nucleus as a liquid drop 3D force distributions are very intuitive: the shear forces are concentrated near the drop's surface
and normal force is positive and constant in the drop's bulk:

\be
s(r) =\gamma^{(3D)} \delta(R_0-r),\quad \frac 23 s(r)+p(r) =\frac{2 \gamma^{(3D)}}{R_0} \theta(R_0-r),
\ee
where $R_0$ is a drop's radius and $\gamma^{(3D)}$ is the surface tension. 
Phenomenologically, $\gamma^{(3D)}\simeq 1.0\  {\rm MeV}/{\rm fm}^2$
for large nuclei obtained from Weizs\"acker mass formula. 
For a large nucleus we do not need
to worry about the relativistic corrections as they are parametrically suppressed by $1/(m R_0)\sim 1/A^{4/3}$.

On the light front the corresponding force distributions are distorted due to motion of the observer and have the following (less intuitive) form:

\be
\mathpzc{S}(x_\perp)=\dfrac{\gamma^{(3D)}}{R_0} \dfrac{x_\perp^2}{\sqrt{R_0^2-x_\perp^2}}\ \theta(R_0-x_\perp), \quad
 \dfrac{1}{2}\mathpzc{S}(x_\perp)+\mathpzc{P}(x_\perp)=\dfrac{\gamma^{(3D)}}{R_0}\sqrt{R_0^2-x_\perp^2}\ \theta(R_0-x_\perp).
\ee
For the 2D pressure in the centre we obtain:

\be
\label{eq:p02d}
\mathpzc{P}(0)=\gamma^{(3D)} \simeq 1.0\ \frac{{\rm MeV}}{{\rm fm}^2},
\ee
which is $A$ independent, i.e. it has qualitatively different dependence on the atomic number in comparison with 3D $p(0)\sim A^{-1/3}$.
The 2D line tension (has dimension of the force) is:

\be
\label{eq:gamma2yadro}
\gamma^{(2D)}=\frac\pi 4\ R_0 \gamma^{(3D)}\simeq 0.88\ A^{1/3}\ \frac{{\rm MeV}}{{\rm fm}}.
\ee
It increases with the atomic number $A$, while the analogous quantity in 3D (surface tension) is independent of $A$.

\section*{\boldmath Light front force distributions in the nucleon as chiral soliton}
\label{Sec-5:Skyrme}

The large $N_c $ argumentation by Witten \cite{Witten:1979kh,Witten:1983tx} justified  that
$N_c$ quarks constituting a baryon can be considered in a mean (non-fluctuating) field that does not change as $N_c\to\infty$.
In this picture  the corresponding mean-field can be considered as a classical one, i.e. as a ``chiral soliton". 

The pioneering calculations of the pressure and shear force distributions in the large $N_c$ nucleon were performed in the chiral quark-soliton model 
\cite{Goeke:2007fp} and in the Skyrme model in Refs.~\cite{Cebulla:2007ei,Goeke:2007fq}. More recently these studies were extended 
to the nucleon in the nuclear matter \cite{Kim:2012ts,Jung:2013bya,Jung:2014jja}, to $\Delta$-baryon \cite{Perevalova:2016dln,Panteleeva:2020ejw,Kim:2020lrs}
and to the charmed baryons in Ref.~\cite{Kim:2020nug}. As all these approaches employ the large $N_c$ picture the relativistic corrections
are parametrically suppressed by $1/N_c$, and all distributions can be obtained directly as functionals of the chiral mean-field.

\begin{figure}[ht]
\begin{minipage}[h]{0.45\linewidth}
\center{\includegraphics[width=1\linewidth]{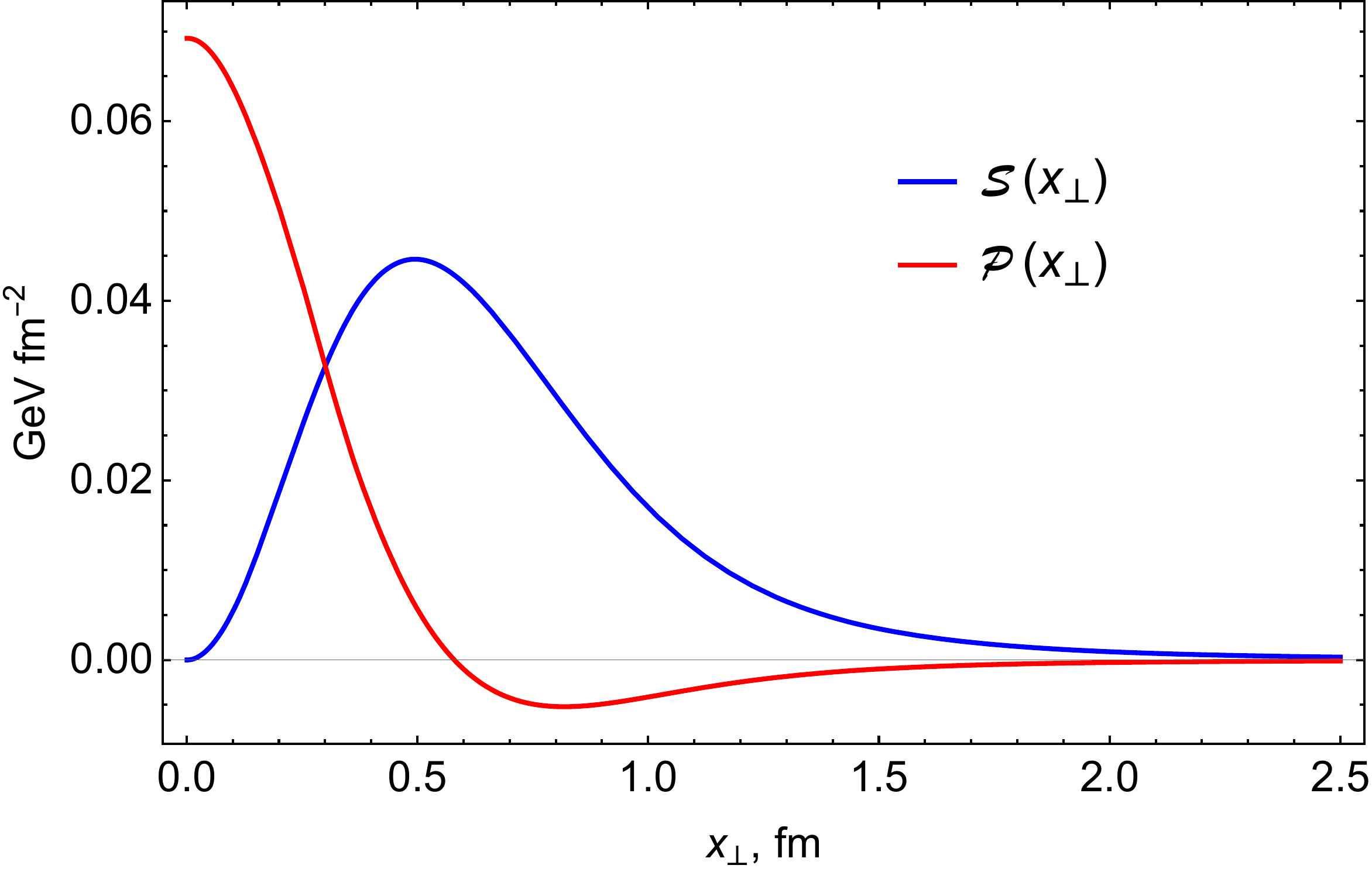}}  
\end{minipage}
\begin{minipage}[h]{0.45\linewidth}
\center{\includegraphics[width=1.025\linewidth]{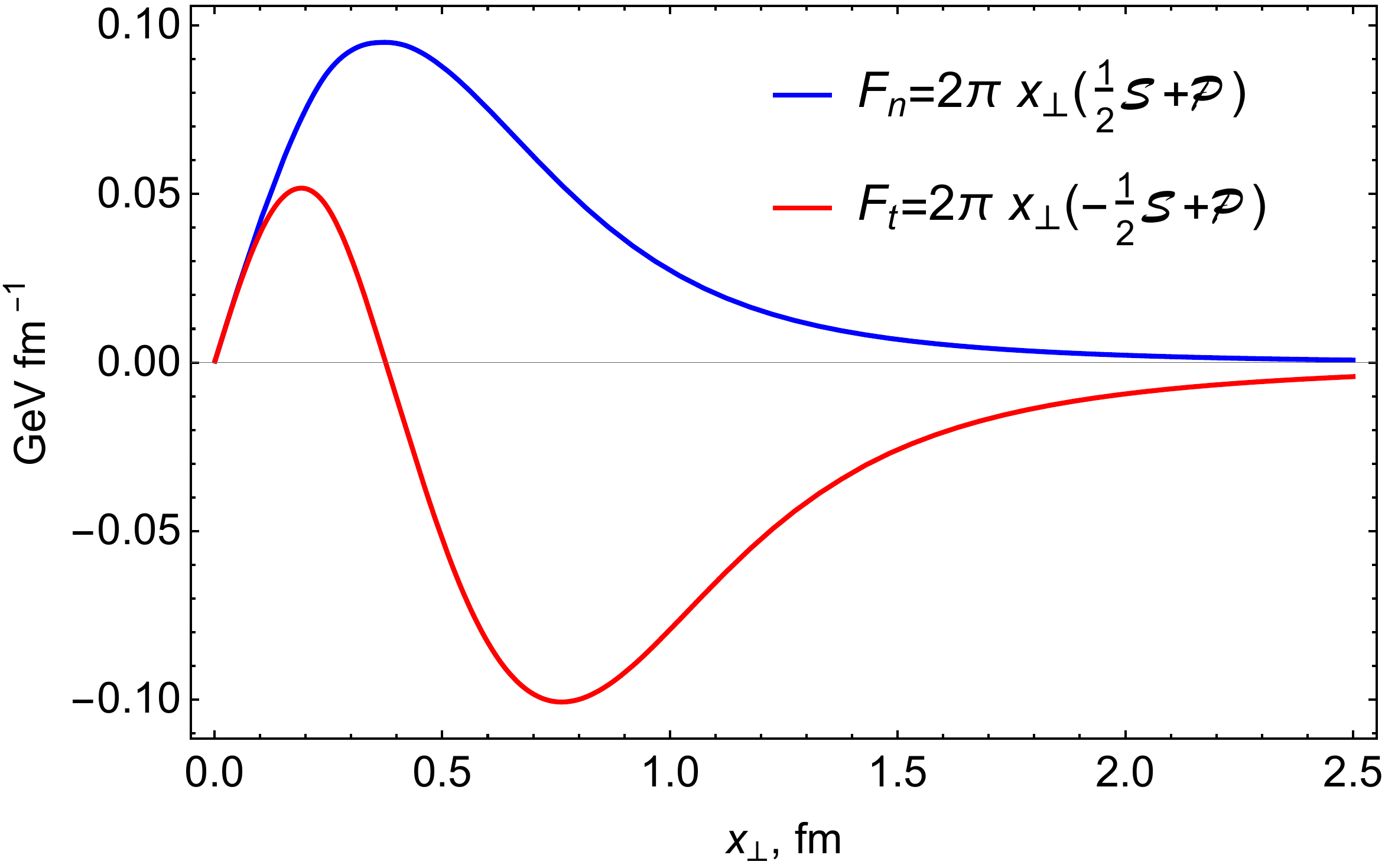}} 
\end{minipage}

\caption{The 2D pressure and share force distributions calculated in Skyrme model (left panel) and distribution of normal
and tangential 2D forces calculated in the same model (right panel). 
}
\label{fig:1}
\end{figure}

We use here the results of Ref.~\cite{Cebulla:2007ei} for 3D $s(r)$ and $p(r)$ to compute  their light front counterparts using 
Eq.~(\ref{eq:p2ds2dr}). The results are shown on Fig~\ref{fig:1} where the light front shear force $\mathpzc{S}(x_\perp)$ and pressure
$\mathpzc{P}(x_\perp)$ are shown in the left panel. On the right panel we show the results for the distribution of normal and transverse forces on light front. 
It is very interesting that  the obtained light front distributions have the shapes very similar to their 3D counterparts (see e.g. Figs.~3,4 of Ref.~\cite{Cebulla:2007ei}).
For example, the transverse force distribution has a node about the same position around $\sim 0.4\div 0.5$~fm in both 3D and 2D, also
both 3D and 2D pressure distributions have nodes at about the same position of $\sim 0.6\div 0.7$~fm. 
 We see that in contrast to the liquid drop the reduction from 3D to 2D does not change the force distributions
 qualitatively. It would be interesting to understand the reason for that.

From the left panel of Fig.~\ref{fig:1} one can easily read off the value of the 2D pressure $\mathpzc{P}(x_\perp)$ in the nucleon centre:

\be
\mathpzc{P}(0)\simeq 70.0\ \frac{{\rm MeV}}{{\rm fm}^2}.
\ee
It is instructive to compare it with the corresponding 2D pressure at the origin for a large nucleus given in (\ref{eq:p02d}).
The line tension also can be easily computed with help of Eq.~(\ref{eq:gamma2}) with the result:
\be
\gamma^{(2D)}\simeq34.0 \ \frac{{\rm MeV}}{{\rm fm}}.
\ee
Again it is instructive to compare it with the result for a large nucleus (\ref{eq:gamma2yadro}).
The 2D mean square mechanical radius is scaled by the geometric factor of 2/3 relative to the 3D mechanical radius.
One can show in the Skyrme model that $\langle x_\perp^2\rangle_{\text{charge}}/\langle r^2\rangle_{\text{charge}}=2/3$.
 Therefore, in 2D we obtain the same
ratio of mechanical radius to charge one as in 3D. The latter was estimated as $\simeq 0.75$ \cite{Polyakov:2018zvc},
hence we obtain:

\be
\dfrac{\langle x_\perp^2\rangle_{\text{mech}}}{\langle x_\perp^2\rangle_{\text{charge}}}\simeq 0.75\quad\quad {\rm [Skyrme\ model]}
\ee

Given the simplicity of the 3D$\rightarrow$2D reduction formula (\ref{eq:p2ds2dr}) the reader can easily obtain any other results
for light front force distributions from the results of Ref. \cite{Cebulla:2007ei}. In general, the Abel transformation relations
(\ref{eq7:abelform},\ref{eq8:abelform}) can used to obtain 2D light front force distributions from any calculation of 3D Breit frame ones.
For example, it is very easy to obtain large $x_\perp$ asymptotic of light front force distributions from the
large $r$ asymptotic of $p(r)$ and $s(r)$ calculated in chiral EFT, see Eq.~(26) of Ref. \cite{Alharazin:2020yjv}.

\section*{Conclusion}

We obtained simple relation  (\ref{eq:p2ds2dr}) between the 2D light front force distributions in a hadron and the 3D Breit frame those.
The expressions for 2D light front force distributions in terms of 3D BF ones were first obtained (in different form) 
and discussed in details in Ref.~\cite{Lorce:2018egm}.
Main new results obtained here are:

\begin{itemize}
\item We showed that the relation between 3D and 2D force distributions is invertible and corresponds to Abel transformation.
 The 3D$\to$2D transformation is given by Eq.~(\ref{eq:p2ds2dr}),
the inverse 2D$\to$3D one is given by Eq.~(\ref{eq:p3ds3dr}). This implies that two types of distributions are mathematically equivalent.
\item We demonstrated that the 3D stability conditions for the force distributions automatically imply the stability of 2D
mechanical system. 
\item
The light front force distributions in the nucleon are computed in the Skyrme model. 
Also these quantities are estimated
for a large nucleus in the liquid drop model of nuclei. Both calculations are important for developing the physics intuition
about the light front force distributions.
\end{itemize}
 It would be interesting to generalise formalism presented here to the case of non-spherically symmetric hadron (with spin $>1/2$).
 We expect that in more general case the Abel transformation is replaced by the Radon one.
 This idea has been addressed but not fully investigated in \cite{Cosyn:2019aio}. 

\acknowledgments

M.V.P.\ is grateful to Jambul Gegelia, C\'edric Lorc\'e, and Peter Schweitzer for 
many illuminating discussions, and to Adam Freese for correspondence.
This work was supported in part by the Deutsche Forschungsgemeinschaft (DFG, German Research 
Foundation)  Project-ID 196253076 TRR 110, and
 by the BMBF (grant 05P2018).

\appendix

\section*{Appendix:. Basics of Abel transformation}

For readers' convenience we give here the basic formulae for Abel transformation.
For the function $g(r)$ its  Abel transform (Abel image) $A[g](x)\equiv\mathpzc{G}(x)$ can be written as:

\be
A[g](x)\equiv\mathpzc{G}(x)=\int\limits_x^\infty \dfrac{dr}{r}\ g(r)\dfrac{1}{\sqrt{r^2-x^2}}.
\ee
The inverse transformation has the form \cite{Abel}:

\be
g(r)=-\frac{2}{\pi}\ r^2 \int\limits_r^\infty dx\ \dfrac{d \mathpzc{G}(x)}{dx}\ \dfrac{1}{\sqrt{x^2-r^2}}.
\ee
We give the Abel images of some basic functions, they can be useful for practical applications, for example:

\be
A\left[\dfrac{1}{r^\alpha}\right](x) =\frac{\sqrt{\pi}}{2}\  \dfrac{\Gamma\left(\frac{\alpha+1}{2}\right)}{ \Gamma\left(
\frac{\alpha+2}{2}\right)}\ \dfrac{1}{x^{\alpha+1}},
\ee
can be used to obtain large $x_\perp$ asymptotic of light front force distributions from the
large $r$ asymptotic of $p(r)$ and $s(r)$ obtained in chiral EFT, see Eq.~(26) of Ref. \cite{Alharazin:2020yjv}.
Further useful Abel images are collected in Table~I.

\begin{table}[h]
\begin{tabular}{c|c}
$g(r)$ & $A[g](x)$\\
\toprule
$r^2 e^{-\lambda r^2}$ & $\dfrac 12 \sqrt{\dfrac{\pi}{\lambda}} e^{-\lambda x^2}$ \\ 
$r e^{-\lambda r}$& $K_0(\lambda x)$\\ 
$r \sin(\omega r)$ & $\dfrac{\pi}{2}J_0(\omega x)$\\ 
$r \cos(\omega r)$ & $-\dfrac{\pi}{2}Y_0(\omega x)$\\ 
$r\ K_0(\lambda r)$  &  $\dfrac 12 K_0\left(\frac{\lambda x}{2}\right)^2$\\
$r\ K_1(\lambda r)$  &  $\dfrac{\pi}{2}\ \dfrac{e^{-\lambda x}}{\lambda x}$\\
$r\ J_1(\omega r)$  &  $ \dfrac{\sin(\omega x)}{\omega x}$\\
$r\ Y_1(\omega r)$  &  $ -\dfrac{\cos(\omega x)}{\omega x}$\\
$\theta(R_0-r)$  & $\dfrac 1x \arccos\left(\dfrac{x}{R_0}\right)\ \theta(R_0-x)$\\ 
$r^2\ \theta(R_0-r)$  & $\sqrt{R_0^2-x^2}\ \theta(R_0-x)$
\end{tabular}
\label{taba}
\caption{Abel transforms of various functions.}
\end{table}

\end{document}